\documentclass{osa-article}
\journal{osac}
\usepackage{graphicx}
\usepackage{amsfonts}
\usepackage{xspace}
\newcommand{\bra}[1]{\ensuremath{\left\langle#1\right|}}
\newcommand{\ket}[1]{\ensuremath{\left|#1\right\rangle}}
\newcommand{\bracket}[2]{\ensuremath{\left\langle#1 \vphantom{#2}\right| \left. #2 \vphantom{#1}\right\rangle}}

\usepackage{amsmath}
\usepackage{mathtools}
\DeclarePairedDelimiter\abs{\lvert}{\rvert}

\usepackage{calc}
\articletype{Research article}

\begin{document}
\title{Two Mode Photon Added Schrödinger Cat States: Nonclassicality and Entanglement}

\author{S Nibedita Swain,\authormark{1,*} Yashovardhan Jha,\authormark{2} and Prasanta K. Panigrahi \authormark{1,**}}

\address{\authormark{1}Department of Physical Sciences, Indian Institute of Science Education and Research Kolkata, Mohanpur-741246, West Bengal, India.\\
\authormark{2}Department of Physics, Indian Institute of Science Education and Research Bhopal, Bhopal Bypass Rd, Bhauri, Madhya Pradesh 462066, India.
}
\email{\authormark{*}nibedita.iiser@gmail.com} 
\email{\authormark{2} yashoovardhanjha@gmail.com}
\email{\authormark{**}pprasanta@iiserkol.ac.in}

\begin{abstract*}
    The concept of photon added two-mode Schr\"{o}dinger cat state in which both modes are independent is introduced, their non-classical properties and entanglement are studied. The introduced states emerge as the eigenstates of $f_1f_2a_1a_2$, where $f_{1}, f_{2}$ are nonlinear functions of the number operator and $a_{1}, a_{2}$ are annihilation operators. We study the evolution of these states under the canonical transformation using the parity operator for the case of standard coherent states of the harmonic oscillator. The non-classical properties of these states are evaluated especially by considering sub-Poissonian photon statistics and photon number distribution. Interestingly, the addition of photons leads to shifting the region in which photon number distribution shows oscillatory behavior. In addition, the entanglement of introduced states has been quantitatively analyzed using concurrence. We observe that the state approaches the maximum entanglement more rapidly after the addition of photons.
 \end {abstract*}
\section{Introduction}
Continuous variable states \cite{gerry1995nonclassical, agarwal1991nonclassical, adesso2014continuous} are very important in the area of quantum optics. Coherent state \cite{sudarshan1963equivalence, glauber1963coherent} is classical, addition of the photons leads to the manifestation of quantum nature which has been theoretically shown by Agarwal et al. \cite{agarwal1991nonclassical} and experimentally verified recently \cite{zavatta2004quantum}. 
In 1995, Gerry and Grobe \cite{gerry1995nonclassical} introduced correlated two-mode Schr\"{o}dinger cat states which are eigenstates of difference of the mode number operators. These states can be created in a variety of ways \cite{zhou2021generating,zapletal2022stabilization}, including atoms trapped in optical cavities \cite{duan2019creating} and by sending {${\Lambda}$}-types of three-level atoms through many cavities filled with coherent fields \cite{zheng1998scheme} and today, these states are easily accessible on different experimental schemes \cite{duan2019creating, chen2021shortcuts, zhou2021generating, zapletal2022stabilization,li2021quantum}.
Cat states have intrinsic quantum nature which have been recently experimentally analyzed with the cat code being in reality to use of continuous variable states for quantum information processing. Addition of a single photon to the cat state and its physical manifestation including statistics have been recently investigated \cite{tyagi2021photon}.

Entangled continuous variable states \cite{nibedita2017generalized, agarwal2012quantum} have been recently investigated with metrology \cite{bhatt2008entanglement} and other applications in quantum teleportation \cite{harris2021quantum}, quantum computation \cite{biamonte2021universal}, quantum key distribution \cite{scarani2009security}.
In a quantum optics research, two-mode cat state is generally used as an entangled tool, which cannot be distilled only by classical communications due to the limitation from the no-go theorem \cite{takahashi2010entanglement}. To suffice the condition of quantum information protocols for long-distance communication, there have been suggestions and realizations for manipulating the quantum states, which are reasonable ways to conditionally operate a nonclassical state \cite{kohnke2021quantum, agarwal2005quantitative, kannan2021positive} of an optical field by subtracting or adding photons from/to a optical states \cite{agarwal1991nonclassical,agarwal2012quantum, tyagi2021photon,ghosh2015mesoscopic}. Actually, the photon addition and subtraction have been successfully demonstrated experimentally for probing quantum commutation rules by Parigi et al. \cite{parigi2007probing}. It would be interest to investigate the effect of photon addition in cat state because of its intrinsic non-local character.
The entangled cat state being non-local, it is worse investigating the effect of addition of a single photon because of its particle nature of the photon is expected to get delocalized, in particular phase-space behaviour is worse investigating.
In this paper, we focus on the effect of photon addition to the two mode entangled Schr\"{o}dinger cat states separated in phase by {$\pi$} in which both modes are independent and the metrology application of these states are being investigated. We evaluate the state obtained by repeated application of different number of photon creation operators on the both modes of the two mode cat states. We derive the evolution of these states under canonical transformation using a parity operator. Such a state has entanglement and is shown to exibit non-classical behaviour. We use photon number distribution \cite{kono2017nonclassical} and Mandel parameter\cite{mandel1979sub} for studying nonclassical behavior of two-mode photon added Schr\"{o}dinger cat states. According to our findings, the photon number distribution is oscillatory, evidence of nonclassical behavior for the introduced states. Interestingly, with the addition of photons to both modes, the region in which photon number distribution shows oscillatory behavior shifts away from the origin. For both modes of given states, the Mandel parameter is also calculated and the results revealed negativity. Negativity in the Mandel parameter is another sign of nonclassical behavior\cite{mandel1979sub}. Finally, we study in detail the quantification of entanglement using concurrence \cite{wootters1998entanglement, kuang2003generation,parit2019correlated}.
Below is a detailed outline of how the paper is organized. 
In section 2, we give an analytical expression for our general photon added two-mode Schrödinger cat states. After that, in the next section, we derive the canonically transformed photon added two mode cat states . Section 4 of this paper presents a study of various nonclassical properties of the introduced states. In section 5, we study the entanglement of these states and quantified it using concurrence.

\section{Generation of Photon Added Two Mode Schr\"{o}dinger Cat States}
In this section we construct photon-added two-mode Schr\"{o}dinger cat states.
We have added photons means two creation operators independently on the first and second mode of the two-mode cat state to get the introduced state.

 The cat states with relative phase $\phi$ ranging from 0 to $\pi$ \cite{zheng1998scheme,anbaraki2018non}
\begin{equation}
\ket{\psi}_c = N(\ket{\alpha_1,\alpha_2,f_1,f_2} + e^{i\phi}\ket{-\alpha_1,-\alpha_2,f_1,f_2})    \nonumber
\end{equation}
where N is the normalization constant. Also ${f_1}$ and ${f_2}$ are operator-valued function of the number operator which we take as ${f_1 = f_2 = 1}$ for our case since we deal with ordinary coherent and cat states. As a result of simplifying the above state, it can be expressed as follows
\begin{equation}
\ket{\psi}_c = N(\ket{\alpha_1,\alpha_2} + e^{i\phi}\ket{-\alpha_1,-\alpha_2})    \nonumber
\end{equation}
It is easy to show that for the independent annihilation operators $\hat{a}$ and $\hat{b}$, the two-mode photon added cat state {$\ket{\psi}_c$} is an eigenstate of operator {$\hat{a}_1\hat{a}_2$}
\begin{equation}
    \hat{a}_1\hat{a}_2\ket{\psi}_c = \alpha_1 \alpha_2 \ket{\psi}_c \nonumber
\end{equation}

The problem arises when it comes to generation of such states. In this paper, we analyse this concern.
Now the photon added two mode Schr\"{o}dinger cat state can be expressed as
\begin{equation}
    \ket{\psi,m_1,m_2}_c = N_3{{a_1^{\dagger}}^{m_1}}{{a_2^{\dagger}}^{m_2}}(\ket{\alpha_1,\alpha_2} + e^{i\phi}\ket{-\alpha_1,-\alpha_2}) \nonumber
\end{equation}
where ${a_1^{\dagger}}$ and ${a_2^{\dagger}}$ are the creation operators for first and second mode and ${N_3}$ is the normalization constant.
   $$ \ket{\psi,m_1,m_2}_c = N_3({{a_1^{\dagger}}^{m_1}}\ket{\alpha_1} \otimes {{a_2^{\dagger}}^{m_2}}\ket{\alpha_2} + e^{i\phi}{{a_1^{\dagger}}^{m_1}}\ket{-\alpha_1} \otimes {{a_2^{\dagger}}^{m_2}}\ket{-\alpha_2}) $$
\\
After Simplification,
\begin{equation}
     \ket{\psi,m_1,m_2}_c = N_3{k_1(\alpha_1,m_1)}^{-1}{k_2(\alpha_2,m_2)}^{-1}(\ket{\alpha_1,m_1} \otimes \ket{\alpha_2,m_2} + e^{i\phi}\ket{-\alpha_1,m_1} \otimes \ket{-\alpha_2,m_2}) \label{eq1}
\end{equation}
$\ket{\alpha,m}$ is normalized photon added coherent states with $k_i(\alpha_i,m_i)$ with $i=1,2$ as normalization constant, where $k_i(\alpha_i,m_i)$ can be expressed using Eq.\eqref{eq1} ${N_3}$ can be calculated as
\begin{equation}
    N_3 = \bigg(2e^{{-\mid\alpha_1\mid^2}-{\mid\alpha_2\mid^2}}\sum_{n_1=0}^{\infty} \sum_{n_2=0}^{\infty}\Big[\frac{{\mid{\alpha_1}\mid}^{2n_1}{\mid{\alpha_2}\mid}^{2n_2}(n_1+m_1)!(n_2+m_2)!(1+\cos{\phi}(-1)^{n_1+n_2})}{(n_1!n_2!)^2}\Big]  \bigg)^{-\frac{1}{2}} \nonumber
\end{equation}
In number state basis, the photon added two-mode Schr\"{o}dinger cat state is
\begin{equation}
\begin{aligned}
    \ket{\psi,m_1,m_2}_c = N_3e^{-\frac{\mid\alpha_{1}\mid^2}{2}-\frac{\mid\alpha_{2}\mid^2}{2}} \sum_{p_1=0}^{\infty} \sum_{p_2=0}^{\infty}\Big[\frac{{\alpha_1^{p_1}}{\alpha_2^{p_2}}\sqrt{(p_1+m_1)!(p_2+m_2)!}(1+e^{i\phi}(-1)^{p_1+p_2})}{p_1!p_2!}\Big] \\ \ket{p_1+m_1,p_2+m_2}
    \end{aligned}
\end{equation}
After adding photons to both modes, these states are nonlinear states which are eigenstates of the operator {$\hat{f}_1\hat{f}_2\hat{a}_1 \hat{a}_2$} 
\begin{equation}
    \hat{f}_1\hat{f}_2\hat{a}_1 \hat{a}_2 \ket{\psi,m_1,m_2}_c = \alpha_1 \alpha_2 \ket{\psi,m_1,m_2}_c \nonumber
\end{equation}
where {$\hat{f}_1$} and {$\hat{f}_2$} can be expressed as \cite{sivakumar1999photon}
\begin{equation}
    \hat{f}_i(\hat{n_i},m_i) = 1 - \frac{m_i}{1 + a_i^{\dagger}a_i} \label{eq2}
\end{equation}
Eq. \eqref{eq2} implies that photon added two-mode cat state can also be considered as nonlinear cat states. These states can be physically realized by the most fundamental process degaussification in which matter links with the light.
\section{Canonical Transformation}

This section describes the canonical transformation of the introduced photon-added two-mode cat states for the case of standard coherent states of a harmonic oscillator using the parity operator. Our objective is to study how the structure of the states changed under this transformation. We will find the transformed states by finding the eigenstates of $\hat{F}_1\hat{F}_2\hat{A}_1\hat{A}_2$, where $\hat{F}_1, \hat{F}_2, \hat{A}_1, \hat{A}_2$ are transformed operators. In order to achieve this, we have used the transformed two-mode coherent states. Under this transformation using parity operator $\hat{P}$, the operators {$\hat{a}_i$}, {$\hat{a}_i^{\dagger}$}, {$\hat{f}_1$} and {$\hat{f}_2$} transformed as\cite{spiridonov2021superpositions}

\begin{equation}
    \hat{A}_i = \hat{P}\hat{a}_i, \; \; \hat{A}_i^{\dagger} = \hat{a}_i^{\dagger}\hat{P} \nonumber
\end{equation}
\begin{equation}
    \hat{F}_i = 1 - \frac{m_i}{1 + A_i^{\dagger}A_i}, \; \;  \hat{F}_i = 1 - \frac{m_i}{1 + a_i^{\dagger}P^2a_i}  \nonumber
\end{equation}

Where $\hat{P}$ is the parity operator which is acting on both modes. Now using {$P^2 = 1$} 
\begin{equation}
    \hat{F}_i = \hat{f}_i  \nonumber
\end{equation}

The transformed two mode coherent states can be expressed as\cite{spiridonov1995universal,spiridonov2021superpositions}

\begin{equation}
    \ket{\alpha_1,\alpha_2}_P = \frac{1}{\sqrt{2}}(e^{\frac{-i\pi}{4}}\ket{i\alpha_1 ,i\alpha_2 } + e^{\frac{i\pi}{4}} \ket{-i\alpha_1 ,-i\alpha_2} )  \nonumber
\end{equation}
\begin{equation}
    \ket{-\alpha_1,-\alpha_2}_P = -\frac{1}{\sqrt{2}}(e^{\frac{-i\pi}{4}}\ket{i\alpha_1 ,i\alpha_2 } + e^{\frac{i\pi}{4}} \ket{-i\alpha_1 ,-i\alpha_2} )  \nonumber
\end{equation}

Consider the transformed two mode cat state is ${\ket{\psi}}_c^P$ then

\begin{equation}
     {\hat{A}_1}{\hat{A}_2}\ket{\psi}_c^P = {\alpha_1}{\alpha_2} \ket{\psi}_c^P   \nonumber
\end{equation}

where $\ket{\psi}_c^P$ can be expressed as

\begin{equation}
    {{\ket{\psi}}_{c}}^{P} = {N_{P}}^{'}(\ket{\alpha_{1},\alpha_{2}}_{P} + e^{i\phi}\ket{-\alpha_{1},-\alpha_{2}}_{P}) \nonumber
\end{equation} 

\begin{equation}
    {\ket{\psi}_c}^P = \frac{{N_P}^{'}}{\sqrt{2}}(e^{\frac{-i\pi}{4}}\ket{i\alpha_1 ,i\alpha_2 } + e^{\frac{i\pi}{4}} \ket{-i\alpha_1 ,-i\alpha_2} ) (1-e^{i\phi})  \nonumber
\end{equation}

where $N_p^{'}$ is the normalization constant. Now consider the transformed photon added cat states are $\ket{\psi,m_1,m_2}_c^P$ then

\begin{equation}
    \ket{\psi,m_1,m_2}_c^P = A_1^{{\dagger}^{m_1}} A_2^{{\dagger}^{m_2}} \ket{\psi}_c^P  \nonumber
\end{equation}

 \begin{equation}
  \begin{aligned}
    \ket{\psi,m_1,m_2}_c^P = \frac{N_P}{\sqrt{2}} (e^{\frac{-i\pi}{4}}\ket{i(-1)^{m_1+m_2}\alpha_1,m_1} \otimes \ket{i(-1)^{m_1+m_2}\alpha_2,m_2} + e^{\frac{i\pi}{4}}\\ \ket{-i(-1)^{m_1+m_2}\alpha_1,m_1} \otimes \ket{-i(-1)^{m_1+m_2}\alpha_2,m_2} ) \times (1-e^{i\phi})   \nonumber
   \end{aligned}
 \end{equation}
     
where $\ket{i(-1)^{m_1+m_2}\alpha_j,m_j}$ and $\ket{-i(-1)^{m_1+m_2}\alpha_j,m_j}$ for $j=1,2$ are the normalized photon added coherent states\cite{karimi2017two}. $N_p$ is the normalization constant. The transformed photon added cat states satisfies
\begin{equation}
  \hat{F}_1\hat{F}_2\hat{A}_1\hat{A}_2 \ket{\psi,m_1,m_2}_c^P = \alpha_1\alpha_2 \ket{\psi,m_1,m_2}_c^P  \nonumber
\end{equation}

So the transformed photon added cat states, are the non-linear cat states with relative phase $\frac{\pi}{2}$ with a factor of $(1-e^{i\phi})$, which depends on the relative phase $\phi$ of the introduced states before transformation. The new states are also eigenstates of $\hat{F}_1\hat{F}_2\hat{A}_1\hat{A}_2$. It is because the transformed system represents the same evolving system.

\section{Nonclassicality}
This section demonstrates non-classical behavior of introduced states using various phenomena such as photon number distribution and sub-Poissonian photon statistics. There have been studies showing that oscillations in photon number distribution and negativity of the Mandel parameter indicate nonclassical behavior \cite{afshar2016nonclassical,gerry1995nonclassical}. Also, the phase space structure of cat state\cite{tyagi2021photon,pan2013cat}, compass state\cite{bhatt2008entanglement} and squeezing properties of various squeezed states has been studied\cite{kannan2021positive}.
If the photon number distribution oscillates then this oscillation is a sure sign of nonclassicality. Photon number distribution ${P(q_1,q_2)}$ is the probability of finding ${q_1}$ photons in first mode and ${q_2}$ photons in second mode which is defined as \cite{roy1998nonclassical,afshar2016nonclassical}

\begin{equation}
    P(q_1,q_2) = \mid\bracket{q_1,q_2}{\psi,m_1,m_2}\mid^2 \label{eq4}
\end{equation}
We obtain
\begin{equation}
\begin{aligned}
    P(q_1,q_2) = \abs[\Bigg]{ N_3e^{-\frac{\mid\alpha_{1}\mid^2}{2}-\frac{\mid\alpha_{2}\mid^2}{2}}\sum_{p_1=0}^{\infty} \sum_{p_2=0}^{\infty}\Big[\frac{{\alpha_1^{p_1}}{\alpha_2^{p_2}}\sqrt{(p_1+m_1)!(p_2+m_2)!}(1+e^{i\phi}(-1)^{p_1+p_2})}{p_1!p_2!}\Big]\\\bracket{q1,q2}{p_1+m_1,p_2+m_2}}^{2} \nonumber
    \end{aligned}
\end{equation}

Now {$${\bracket{q1,q2}{p_1+m_1,p_2+m_2}=1}$$}  
when {${q_1 = p_1+m_1}$ and ${q_2=p_2+m_2}$} which implies that when {${p_1 = p_2 =0}$ then ${q_1 = m_1 }$} and {${ q_2 = m_2}$} so values of {${q_1}$} and {${q_2}$} will start from {${m_1}$} and {${m_2}$} respectively. Now we can write the above expression for photon number distribution as

\begin{equation}
    P(q_1,q_2) =\abs[\Bigg]{ N_3e^{-\frac{\mid\alpha_{1}\mid^2}{2}-\frac{\mid\alpha_{2}\mid^2}{2}}\frac{{\alpha_1^{q_1-m_1}}{\alpha_2^{q_2-m_2}}\sqrt{(q_1)!(q_2)!}(1+e^{i\phi}(-1)^{q_1+q_2-m_1-m_2})}{(q_1-m_1)!(q_2-m_2)!}}^{2} \nonumber
\end{equation}

\begin{figure}[ht]
    \centering
    \includegraphics[width=9cm]{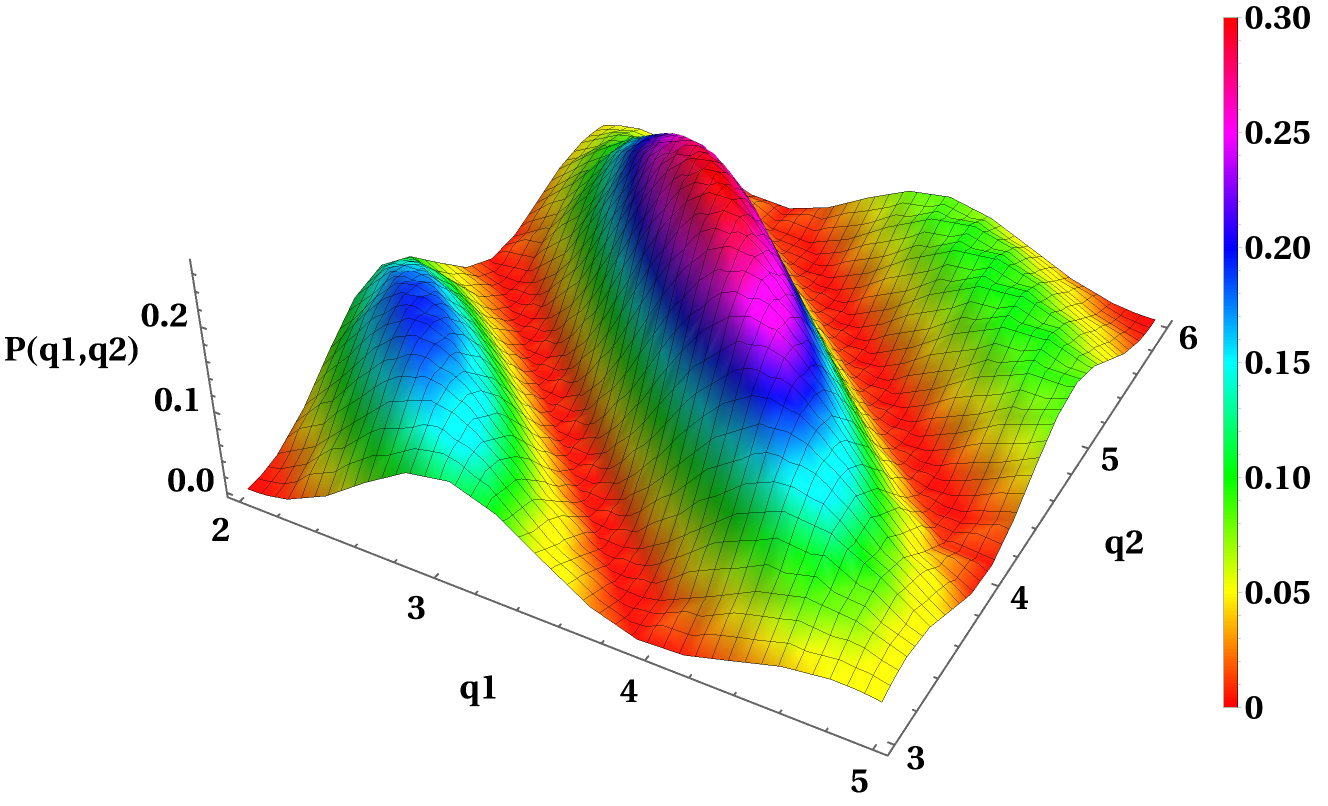}
    \caption{photon number distribution ${P(q_1,q_2)}$ as a function of ${q_1}$ and ${q_2}$ with ${m_1}=2, m_2=3, |\alpha_1|=0.9, |\alpha_2|=0.8, \phi=\pi $}
    \label{fig:photon number dist}
\end{figure}
\begin{figure}[ht]
    \centering
    \includegraphics[width=13cm]{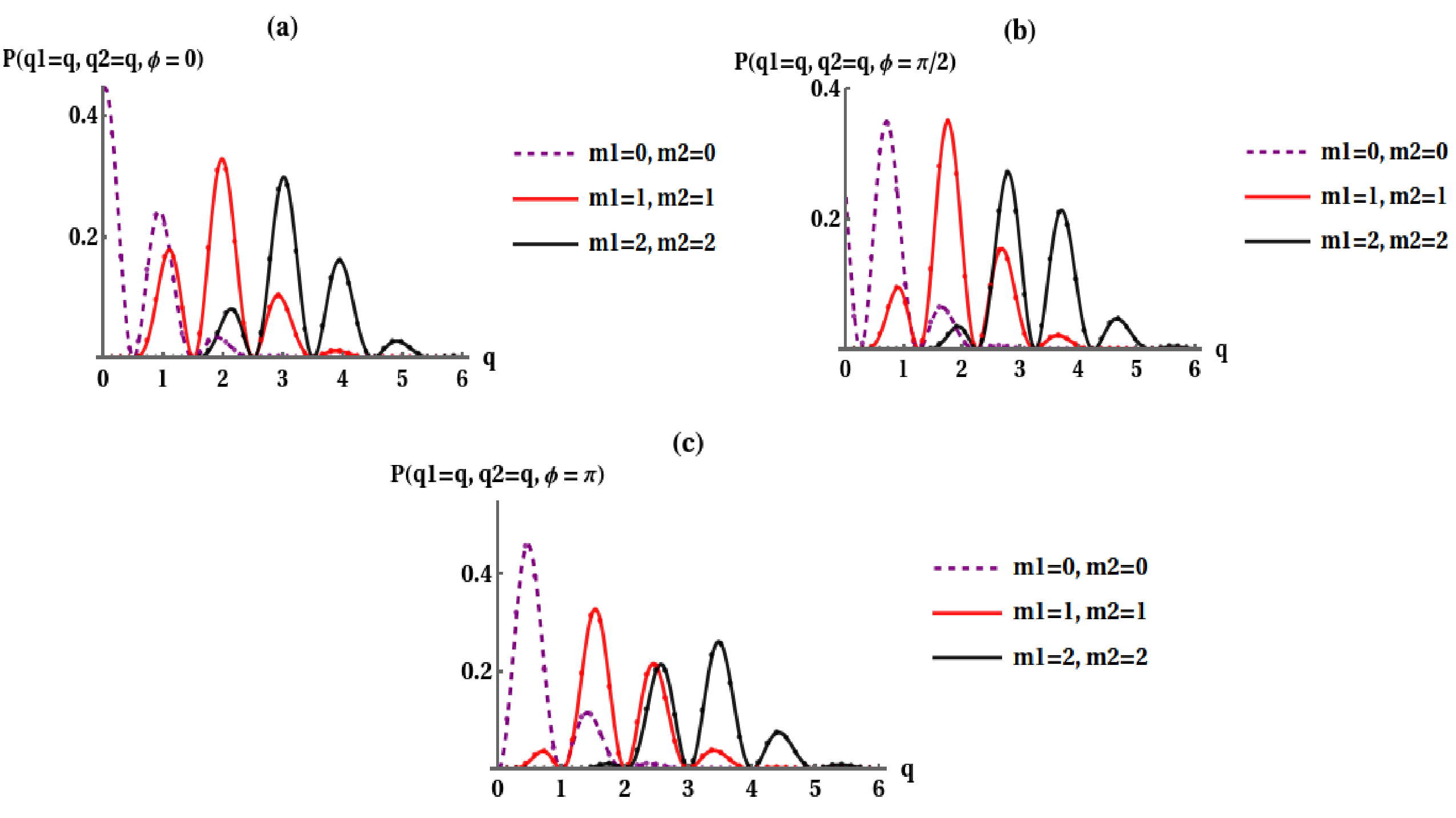}
    \caption{Addition of photons to introduced states resulting in shifting of photon number distribution away from origin (a) photon number distribution ${P(q_1=q,q_2=q,\phi=0)}$ as a function of ${q}$ with {$|\alpha_1|=0.9, |\alpha_2|=0.8$}, (b) photon number distribution ${P(q_1=q,q_2=q,\phi=\frac{\pi}{2})}$ as a function of ${q}$ with {$|\alpha_1|=0.9, |\alpha_2|=0.8$}, (c) photon number distribution ${P(q_1=q,q_2=q,\phi={\pi})}$ as a function of ${q}$ with {$|\alpha_1|=0.9, |\alpha_2|=0.8$}}
\end{figure}

In fig.1, the photon number distribution shows the oscillatory behaviour and also this distribution falls towards zero as the ${q_1}$ and ${q_2}$ increases which is a sign of nonclassicality. According to fig.2, the region in which the photon number distribution exhibits oscillatory behavior is shifting away from the origin for {$\phi=0, \frac{\pi}{2}$} and {$\pi$}.\\
Let us now consider sub-Poissonian photon statistics of given states, which is another way to prove the nonclassical nature of these states. In order to understand the sub-Poissonian photon statistics we first have to calculate the Mandel parameter of each mode, which is defined as \cite{afshar2016nonclassical}

\begin{equation}
    Q_i = \frac{\left\langle {a_i}^{\dagger2} {a_i}^2 \right\rangle  - \left\langle {a_i}^{\dagger} {a_i} \right\rangle^2}{\left\langle {a_i}^{\dagger} {a_i} \right\rangle} \nonumber
\end{equation}

 where i=1,2

To calculate the expectation values of ${a_i}^{\dagger} {a_i}$ and ${{a_i}^{\dagger 2}} {{a_i}}^2$, we use two-mode photon added cat states as
\begin{equation}
\begin{aligned}
    \ket{\psi,m_1,m_2} = N_3e^{-\frac{\mid\alpha_{1}\mid^2}{2}-\frac{\mid\alpha_{2}\mid^2}{2}} \sum_{p_1=0}^{\infty} \sum_{p_2=0}^{\infty}\Big[\frac{{\alpha_1^{p_1}}{\alpha_2^{p_2}}\sqrt{(p_1+m_1)!(p_2+m_2)!}(1+e^{i\phi}(-1)^{p_1+p_2})}{p_1!p_2!}\Big]\\
    \ket{p_1+m_1,p_2+m_2} \nonumber
    \end{aligned}
\end{equation}

For the calculation of $Q_{i}$
\begin{equation}
    \left\langle {a_i}^{\dagger} {a_i} \right\rangle = \left\langle n_i \right\rangle \nonumber
\end{equation}

\begin{equation}
    \left\langle n_i \right\rangle = \bra{\psi,m_1,m_2} n_i \ket{\psi,m_1,m_2} \nonumber
\end{equation}
Now  {$\left\langle n_i \right\rangle$} becomes
\begin{equation*}
     \begin{aligned}
    \left\langle n_i \right\rangle = {2(N_3e^{-\frac{\mid\alpha_{1}\mid^2}{2}-\frac{\mid\alpha_{2}\mid^2}{2}})}^{2} \sum_{p_1=0}^{\infty} \sum_{p_2=0}^{\infty}  \Big[\frac{\mid\alpha_1\mid^{2p_1}\mid\alpha_2\mid^{2p_2}{(p_1+m_1)!(p_2+m_2)!}(p_i + m_i)}{ (p_1!p_2!)^2 }\Big]\\(1+(-1)^{p_1+p_2}\cos{\phi}) \nonumber
    \end{aligned}
\end{equation*}


\begin{equation}
\begin{aligned}
    \left\langle a_i^{\dagger}a_i \right\rangle = {2(N_3e^{-\frac{\mid\alpha_{1}\mid^2}{2}-\frac{\mid\alpha_{2}\mid^2}{2}})}^{2} \sum_{p_1=0}^{\infty} \sum_{p_2=0}^{\infty}  \Big[\frac{\mid\alpha_1\mid^{2p_1}\mid\alpha_2\mid^{2p_2}{(p_1+m_1)!(p_2+m_2)!} (p_i + m_i)}{ (p_1!p_2!)^2 }\Big]\\(1+(-1)^{p_1+p_2}\cos{\phi}) \nonumber
    \end{aligned}
\end{equation}

Similarly  {$\left\langle {a_i}^{\dagger2} {a_i}^2 \right\rangle$} can be calculated as

\begin{equation}
\left\langle {a_i}^{\dagger2}{a_i}^2 \right\rangle  = \bra{\psi,m_1,m_2} {a_i}^{\dagger2}{a_i}^2 \ket{\psi,m_1,m_2} \nonumber
\end{equation}

$$
    \begin{aligned}
 \left\langle {a_i}^{\dagger2}{a_i}^2 \right\rangle =  {2(N_3e^{-\frac{\mid\alpha_{1}\mid^2}{2}-\frac{\mid\alpha_{2}\mid^2}{2}})}^{2} \sum_{p_1=0}^{\infty} \sum_{p_2=0}^{\infty}  [\frac{\mid\alpha_1\mid^{2p_1}\mid\alpha_2\mid^{2p_2}{(p_1+m_1)!(p_2+m_2)!} }
{ (p_1!p_2!)^2 }]\\(p_i + m_i)(p_i+m_i-1)(1+(-1)^{p_1+p_2}\cos{\phi})  \nonumber   
    \end{aligned} $$





 \begin{figure}[ht]
 \begin{minipage}[b]{0.45\linewidth}
 \centering
 \includegraphics[width=\textwidth]{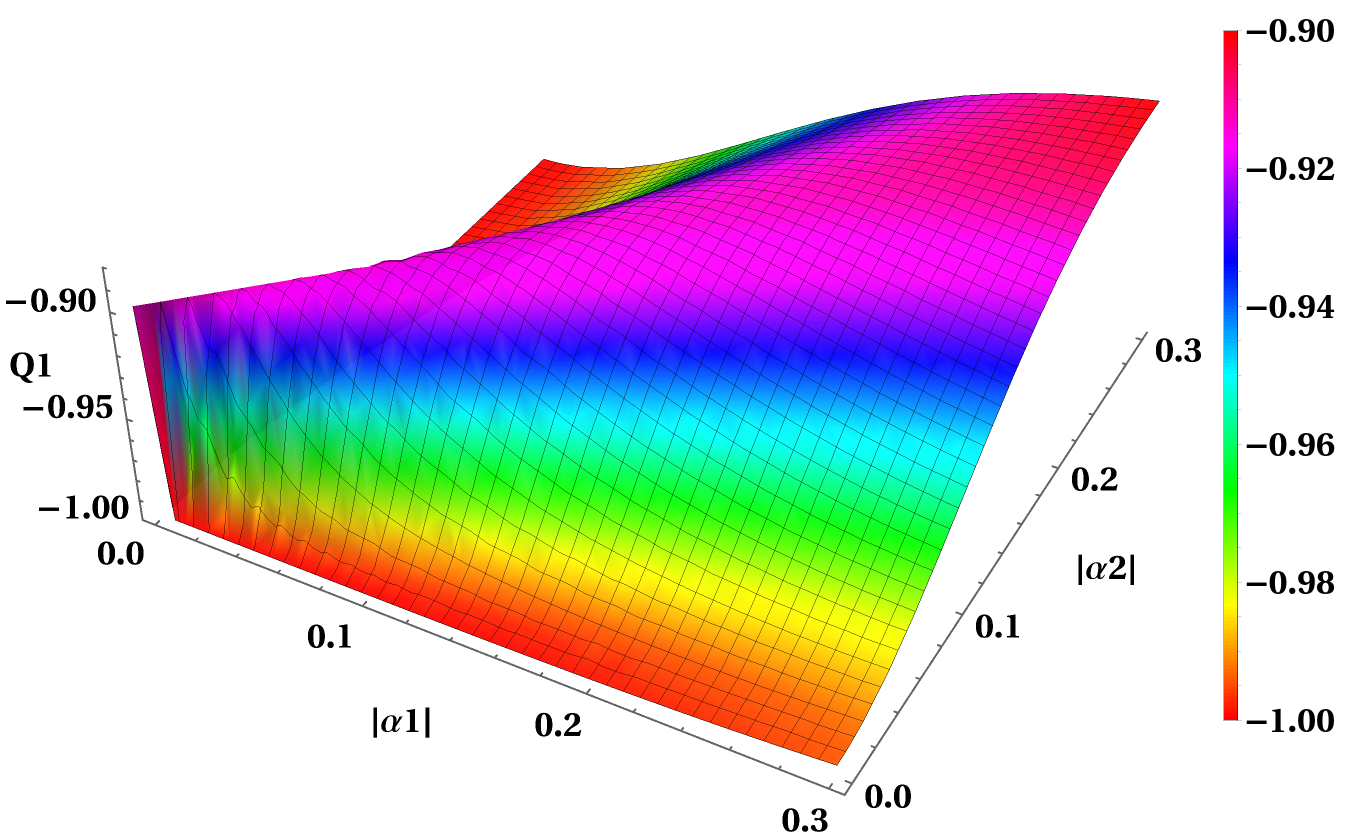}
 \captionsetup{width=0.8\linewidth}
 \caption{3D plot for Mandel parameter for first mode with respect to ${\mid\alpha_1\mid}$ and ${\mid\alpha_2\mid}$ for ${\phi}=\pi$ , ${m_1}=2$ and ${m_2}=3$}
 \label{fig:q1}
 \end{minipage}
 \hspace{0.45cm}
 \begin{minipage}[b]{0.45\linewidth}
 \centering
 \includegraphics[width=\textwidth]{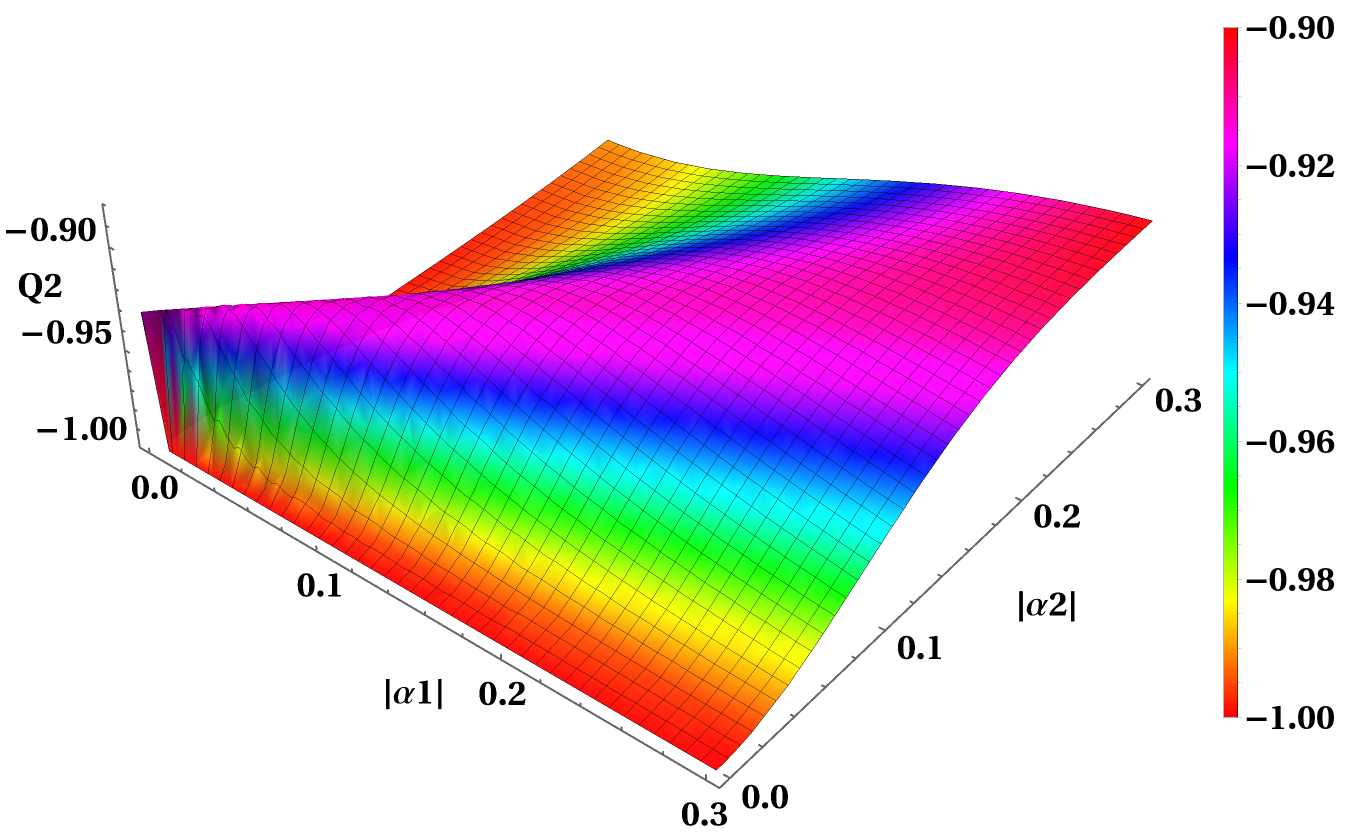}
 \captionsetup{width=0.8\linewidth}
 \caption{3D plot for Mandel parameter for second mode with respect to ${\mid\alpha_1\mid}$ and ${\mid\alpha_2\mid}$ for ${\phi}=\pi$ , ${m_1}=2$ and ${m_2}=3$}
 \label{fig:q2}
 \end{minipage}
 \end{figure}

It is clear from Fig.3 and Fig.4 that Mandel parameter for each mode is showing negativity with respect to {$|\alpha_1|$} and {$|\alpha_2|$} in range from 0 to 0.3 which implies that the state is non-classical in nature. We know that negativity of Mandel parameter shows the sub-Poissonian statistics of the given states which is a pure sign to show the non-classical behaviour of given states. The sub-Poissonian statistics presents photon antibunching in which light interrelate nonlinear medium. In this case of sub-Poissonian statistics, the probability distribution is narrower than Poissonian statistics.

\section{Entanglement}
The goal of this section is to study and quantify the entanglement of photon added two-mode Schrödinger cat states. 
As a quantum correlation\cite{kohnke2021quantum}, entanglement plays a pivotal role in quantum information and computation because of its application in quantum teleportation and computation.
Earlier studies have studied entanglement for various types of one mode states and two-mode states\cite{agarwal2005quantitative}.
In the present section, we study it for more generalized photon-added two-mode Schrödinger cat states, which can be described as the superposition of two modes coherent states. There are various ways to measure entanglement\cite{wootters1998entanglement,nibedita2017generalized}. We calculate the Concurrence for the introduced states to measure the entanglement. Concurrence is a widely used measure to quantify the entanglement in the case of bipartite systems. Concurrence ranges from 0 to 1 (for maximum entangled state). We calculate the Concurrence for the general bipartite entangled states in the beginning, and then using the final expression, we analyse it further for the introduced states.\\
Let's consider a general bipartite entangled state defined as
\begin{equation}
    \ket{\Psi} = M(\lambda \ket{a}\otimes\ket{b} + \delta \ket{c}\otimes\ket{d})
\end{equation}
where ${\lambda}$ and ${\delta}$ are the complex numbers. ${\ket{a}}$ and ${\ket{c}}$ (${\ket{b}}$ and ${\ket{d}}$) are the normalized states of first mode(second mode) with ${\bracket{a}{c}}$ and ${\bracket{b}{d}}$ are non-zero. M is the normalization constant.

The concurrence for the states of the form (4) is defined as
\begin{equation}
    C = \mid 2 M^2 \lambda\delta \sqrt{1-{\mid P_1 \mid}^2}\sqrt{1-{\mid P_2 \mid}^2} \mid
\end{equation}
where ${P_1}$ and ${P_2}$ are given by
\begin{equation}
    P_1 = \bracket{a}{c},
    P_2 = \bracket{d}{b} \nonumber
\end{equation}
To calculate concurrence of the photon added two mode Schrödinger cat states, we can write ${P_1}$,${P_2}$,${M}$,${\lambda}$, and ${\delta}$ by comparing Eqs. \eqref{eq1} and \eqref{eq4}
\begin{equation}
    M = N_3 k_1(\alpha_1,m_1)^{-1} k_2(\alpha_2,m_2)^{-1}, \; \; \lambda=1, \delta=e^{i\phi} \nonumber
\end{equation}

\begin{equation}
    P_1 = \bracket{\alpha_1, m_1}{-\alpha_1,m_1}, \; \; P_2 = \bracket{\alpha_2,m_2}{-\alpha_2,m_2} \nonumber
  \end{equation}

To evaluate ${P_1}$ and ${P_2}$ we will first calculate ${\bracket{\alpha,m}{-\alpha,m}}$. Which can be obtained as 
\begin{equation}
    {\bracket{\alpha,m}{-\alpha,m}} = k^2(\alpha,m) e^{-{\mid\alpha\mid}^2}\sum_{p=0}^{\infty}{\frac{(-1)^p (p+m)! {\mid\alpha\mid}^{2p}}{p!^2}} \nonumber
\end{equation}
where ${k(\alpha,m)}$ is given by
\begin{equation}
    k(\alpha,m) = \bigg(e^{-\mid\alpha\mid^2}\sum_{p=0}^{\infty}\frac{(p+m)!}{p!^2}\mid\alpha\mid^{2p}\bigg)^{-\frac{1}{2}}
\end{equation}
We can write ${P_1}$ and ${P_2}$ as
\begin{equation}
    P_1 = k_1^2(\alpha_1,m_1)e^{-{\mid\alpha_1\mid}^2}\sum_{p=0}^{\infty}{\frac{(-1)^p (p+m_1)! {\mid\alpha_1\mid}^{2p}}{p!^2}} \nonumber
\end{equation}

\begin{equation}
    P_{2} = k_{2}^{2}(\alpha_{2},m_{2})e^{-|\alpha_{2}|2}\sum_{p=0}^{\infty}{\frac{(-1)^p (p+m_{2})! |\alpha_{2}|^{2p}}{p!^{2}}} \nonumber
\end{equation}

Now concurrence for photon added two mode Schr\"{o}dinger cat states can be written as
\begin{equation}
    C = 2\mid {(N_3 k_1(\alpha_1,m_1)^{-1} k_2(\alpha_2,m_2)^{-1})^2 \sqrt{1-{\mid P_1 \mid}^2}\sqrt{1-{\mid P_2 \mid}^2}}\mid
\end{equation}

For the purpose of examining how the concurrence changes with respect to different variables, we will plot the concurrence for photon added two mode Schrödinger cat states with respect to ${\mid \alpha_1 \mid}$ and/or ${\mid \alpha_2\mid}$ for different values of ${m_1}$ and ${m_2}$ with ${\phi} = \pi$.

\begin{figure}[ht]
    \centering
    \includegraphics[width=8cm]{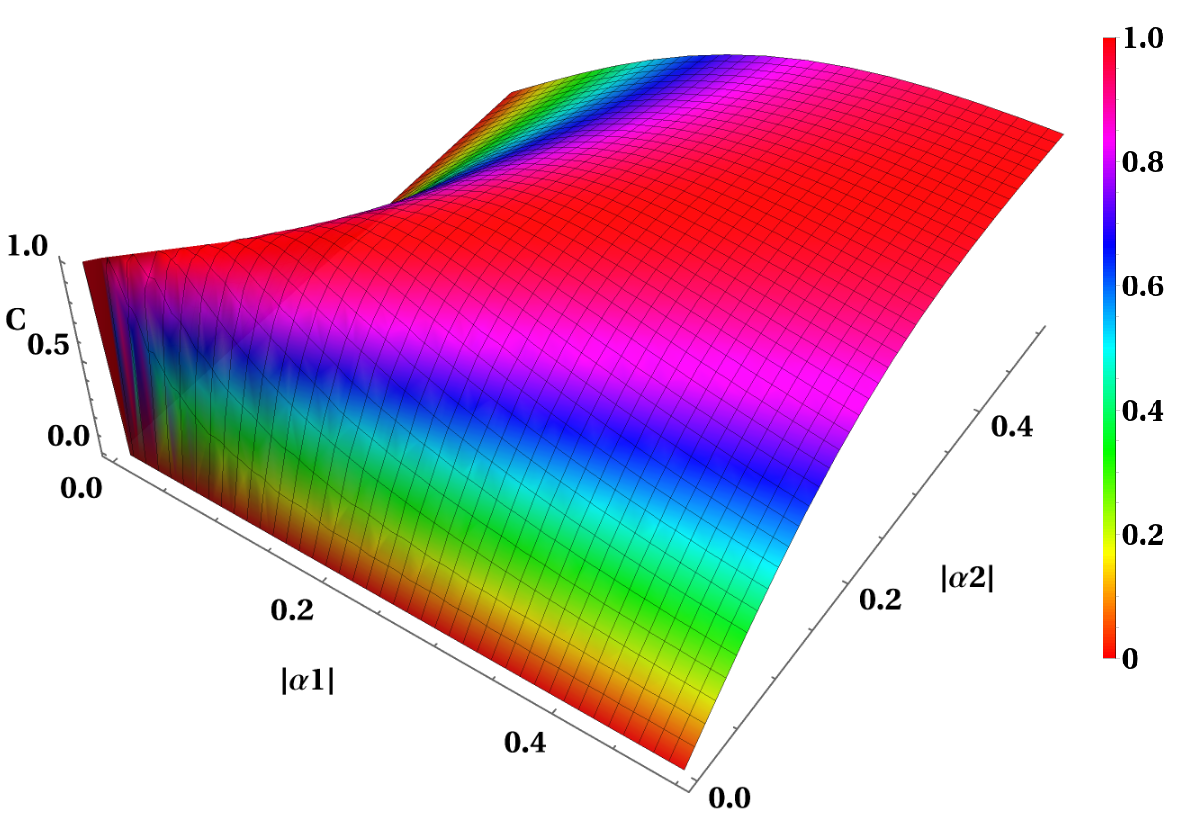}
    \caption{3D plot for concurrence of Photon added two mode Schr\"{o}dinger cat states with respect to ${\mid\alpha_1\mid}$ and ${\mid\alpha_2\mid}$ with ${\phi}=\pi$ , ${m_1}=2$ and ${m_2}=3$ }
    \label{fig:concurrence plot}
\end{figure}




 \begin{figure}[ht]
 \begin{minipage}[b]{0.45\linewidth}
 \centering
 \includegraphics[width=\textwidth]{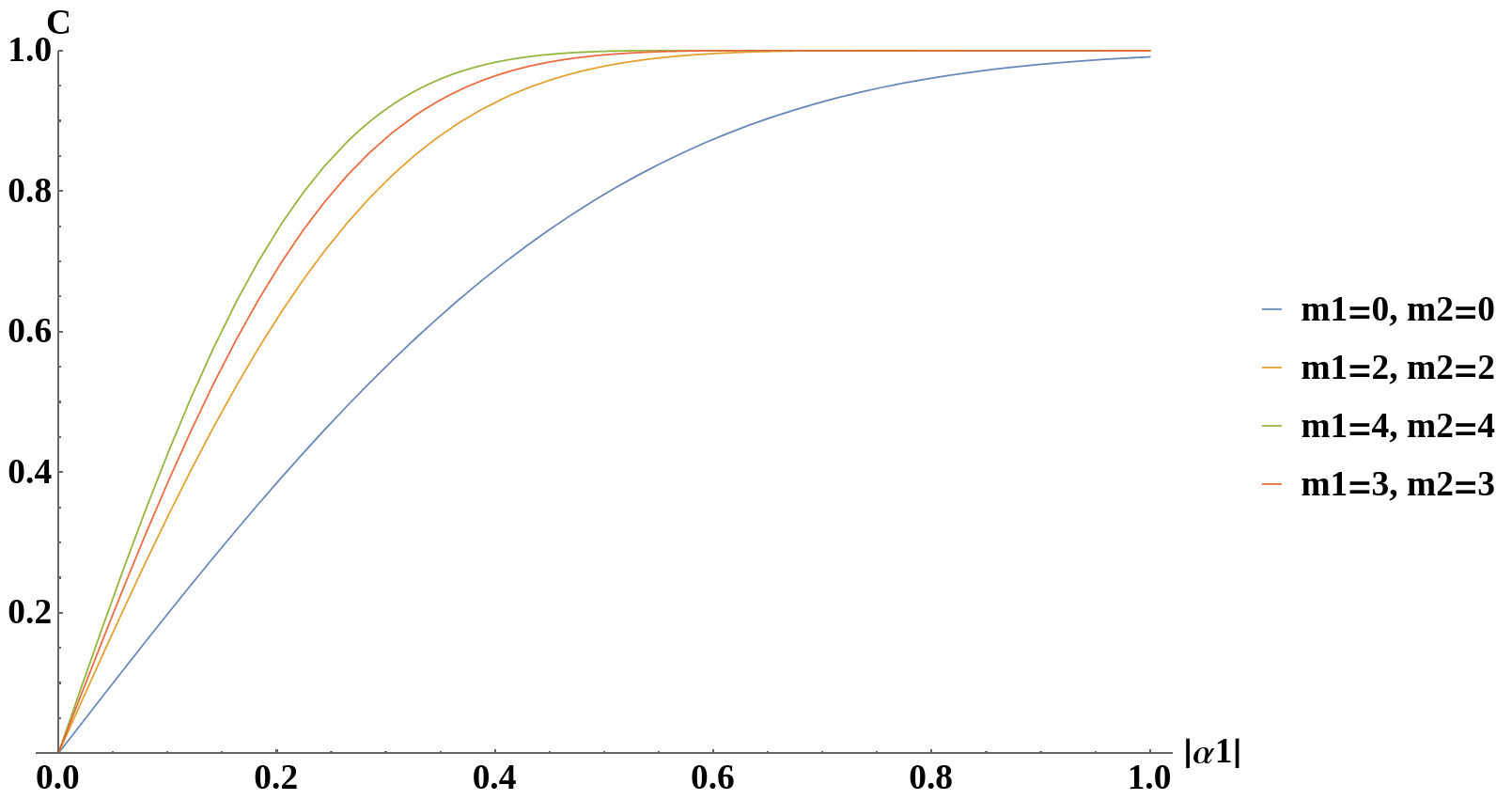}
 \captionsetup{width=0.8\linewidth}
 \caption{Concurrence plot of Photon added two mode Schrödinger cat states with respect to ${\mid \alpha_1 \mid}$ for ${\mid\alpha_2\mid}$=2 and ${\phi}=\pi$ with different values of ${m_1}$ and ${m_2}$}
 \label{c_patmscs}
 \end{minipage}
 \hspace{0.45cm}
 \begin{minipage}[b]{0.45\linewidth}
 \centering
 \includegraphics[width=\textwidth]{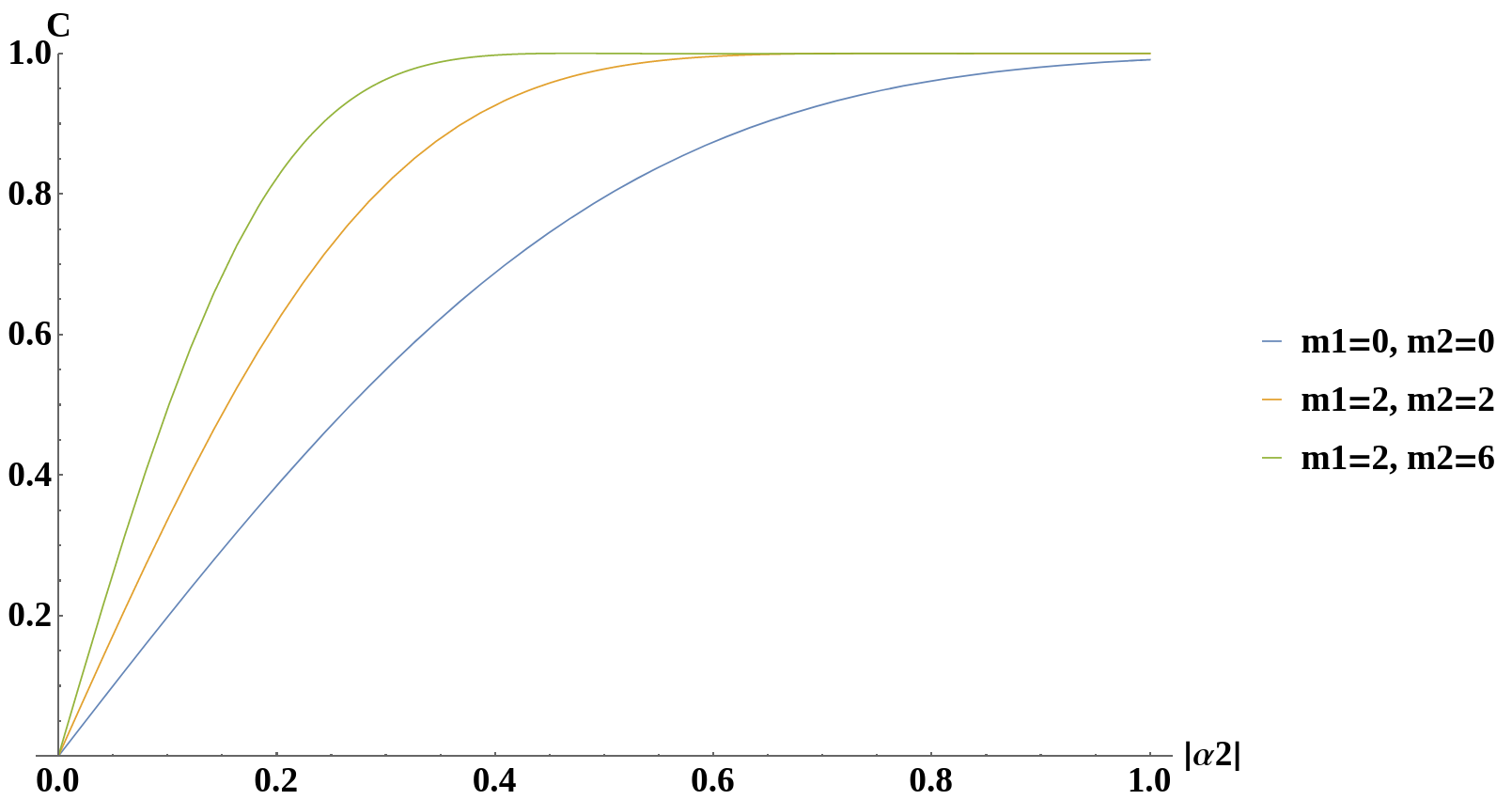}
 \captionsetup{width=0.8\linewidth}
 \caption{Concurrence plot of Photon added two mode Schrödinger cat states with respect to ${\mid\alpha_2\mid}$ for ${\mid\alpha_1\mid}$=2 and ${\phi}=\pi$ with different values of ${m_1}$ and ${m_2}$}
 \label{cpatmscs}
 \end{minipage}
 \end{figure}


\begin{figure}[ht]
    \centering
    \includegraphics[width=13cm]{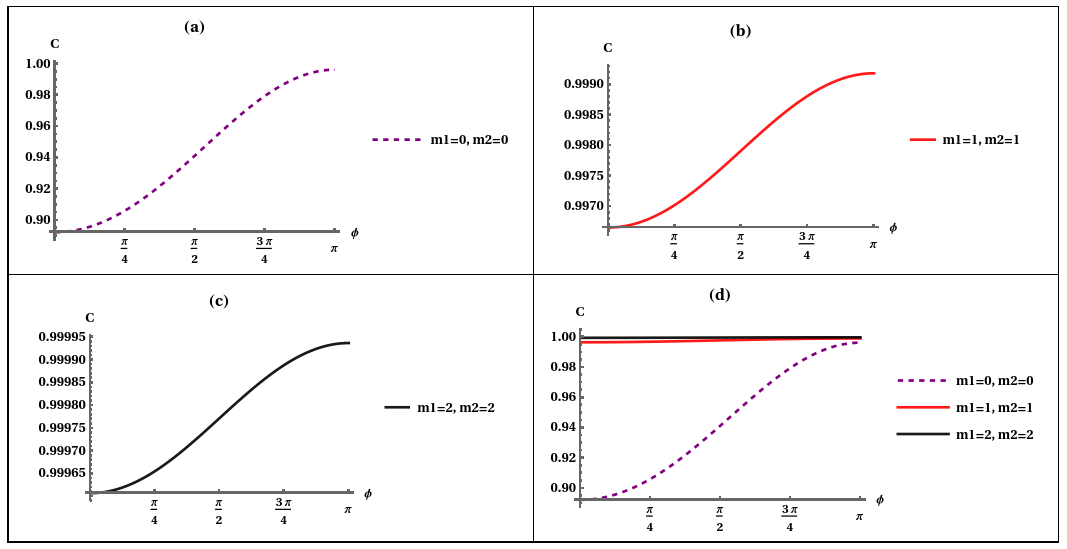}
    \caption{ Concurrence plot of Photon added two mode Schrödinger cat states with  ${\mid\alpha_1\mid}$ = 0.9 for ${\mid\alpha_2\mid}$= 0.8 with respect to ${\phi}$ for different values of ${m_1}$ and ${m_2}$}
\end{figure}
\begin{figure}[ht]
    \centering
    \includegraphics[width=10cm]{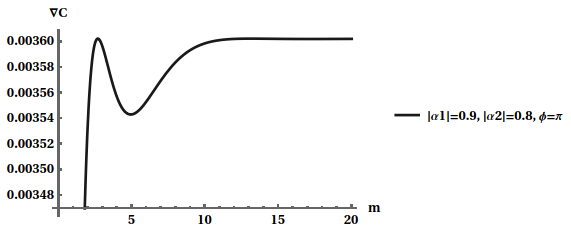}
    \caption{Entanglement difference with respect to added photons for $|\alpha_1| = 0.9$, $|\alpha_2| = 0.8$, and $\phi = \pi$.}
    \label{fig:entanglement difference plot}
\end{figure}
Figures 5, 6, and 7 illustrate that the concurrence for the photon added two-mode Schrödinger cat states starts at zero and is tending to 1(maximum entangled state) as the ${\mid\alpha_1\mid}$ and ${\mid\alpha_2\mid}$ are increasing. It tends to 1 more rapidly as we increase number of added photons ${m_1}$ and ${m_2}$ in both modes.

Furthermore, we calculated the entanglement difference to check how entanglement is changing as more and more photons are being added to both modes of introduced states. Entanglement difference can then be calculated using concurrence. The entanglement difference $ \Delta C$ is calculated as follows:
\begin{equation}
    \Delta C = C(\mid\alpha_{1}\mid, \mid\alpha_{2}\mid, m_{1}, m_{2}, \phi) - C(\mid\alpha_{1}\mid, \mid\alpha_{2}\mid, m_{1}=0, m_{2}=0, \phi) \nonumber 
\end{equation}

where C is the expression of concurrence for introduced states. According to the expression of $\Delta C$, the entanglement difference is the difference between the amount of entanglement before and after photon addition. We have plotted the entanglement difference $\Delta C$ defined as $C(|\alpha_1| = 0.9, |\alpha_2|=0.8, m_1=m, m_2=m, \pi) - C(|\alpha_1|=0.9, |\alpha_2|=0.8, m_1=0, m_2=0, \pi)$ with respect to m in figure 9. According to figure 9, the entanglement difference first increases as the number of photons in both modes increases, then decreases, then again increases and saturates for more number of photons.
\section{Conclusion}
In conclusion, we presented an analytical formulation for the photon-added two-mode {Schr\"{o}dinger} cat states. The given states have been introduced as the superposition of two distinct, coherent states separated in phase by {$\pi$}. Evolution of these states under canonical transformation using parity operator has been studied. We have shown that photon number distribution shows oscillatory behavior, and the Mandel parameter for both modes shows negativity which indicates that these states behave non-classically. The addition of photons to both modes is shifting the photon number distribution away from the origin,  this is an engrossing result. To analyze the entanglement, concurrence is calculated, which is approaching 1 more rapidly, as the added number of photons {$m_1$} and {$m_2$} are increasing. Concurrence is approaching one for every value of phase {$\phi$} as the added number of photons increases. Also, we observed that entanglement difference initially increases, then dips, then increases, and finally saturates as the number of added photons increases in both modes. It would be interesting to study the phase-space structure of these states using the wigner function in four dimensions, why the entanglement difference dips after increasing, and teleportation fidelity using introduced states.
\section{Acknowledgements}
SNS and YJ equally contributed to this work.
SNS is thankful to the University Grants Commission and Council of Scientific and Industrial Research, New Delhi, Government of India for Junior Research Fellowship at IISER Kolkata. PKP acknowledge the partial support from DST-ITPAR grant, India through Grant No. IMT/Italy/ITPAR-IV/QP/2018/G.
\bibliography{sample}
\end{document}